\pdfoutput=1
\documentclass{aa}
\usepackage{natbib}
\usepackage{natbib}
\usepackage{txfonts}
\bibpunct{(}{)}{;}{a}{}{,}
\usepackage[figuresright]{rotating}
\usepackage{savesym}
\usepackage{amsmath}
\usepackage{amsfonts}
\usepackage{color}
\usepackage{graphicx}
\usepackage{appendix}
\usepackage{verbatim}   % useful for program listings
\usepackage{multirow}
\usepackage{subfigure}  % use for side-by-side figures
\usepackage{natbib,twoopt}
\usepackage{morefloats}
\usepackage[breaklinks=true]{hyperref} %% to avoid \citeads line fills
\bibpunct{(}{)}{;}{a}{}{,} %% natbib format for A&A and ApJ
\makeatletter
\newcommandtwoopt{\citeads}[3][][]{\href{http://adsabs.harvard.edu/abs/#3}%
{\def\hyper@linkstart##1##2{}%
\let\hyper@linkend\@empty\citealp[#1][#2]{#3}}}
\newcommandtwoopt{\citepads}[3][][]{\href{http://adsabs.harvard.edu/abs/#3}%
{\def\hyper@linkstart##1##2{}%
\let\hyper@linkend\@empty\citep[#1][#2]{#3}}}
\newcommandtwoopt{\citetads}[3][][]{\href{http://adsabs.harvard.edu/abs/#3}%
{\def\hyper@linkstart##1##2{}%
\let\hyper@linkend\@empty\citet[#1][#2]{#3}}}
\newcommandtwoopt{\citeyearads}[3][][]%
{\href{http://adsabs.harvard.edu/abs/#3}
{\def\hyper@linkstart##1##2{}%
\let\hyper@linkend\@empty\citeyear[#1][#2]{#3}}}
\makeatother
\begin{document}

%% LaTeX will automatically break titles if they run longer than
%% one line. However, you may use \\ to force a line break if
%% you desire.

\title{Dust polarisation studies on MHD simulations of molecular clouds: \\ methods comparison for the relative orientations analysis}

%% You can use \email to mark an email address
%% anywhere in the paper, not just in the front matter.
%% As in the title, use \\ to force line breaks.

\author{Elisabetta R. Micelotta,\inst{1} Mika Juvela,\inst{1} Paolo Padoan,\inst{2,3} Isabelle Ristorcelli,\inst{4,5} Dana Alina\inst{6} \and Johanna Malinen\inst{7}}
\offprints{E. R. Micelotta}
\institute{Department of Physics, PO Box 64, 00014 University of Helsinki,
Finland \\
\email{elisabetta.micelotta@helsinki.fi}
\and
Institut de Ci\`encies del Cosmos, Universitat de Barcelona, IEEC-UB, Mart\'ii Franqu\`es 1, E-08028 Barcelona, Spain
\and
ICREA, Pg. Llu\'is Companys 23, E-08010 Barcelona, Spain
\and
Universit\'e de Toulouse, UPS-OMP, IRAP, F-31028 Toulouse cedex 4, France
\and
CNRS, IRAP, 9 Av. colonel Roche, BP 44346, F-31028 Toulouse cedex 4, France
\and
Department of Physics, School of Science and Technology, Nazarbayev University, Astana 010000, Kazakhstan
\and
Institute of Physics I, University of Cologne, Cologne, Germany
}

\date{Received XX XXX XXXX; accepted XX XXX XXXX}

\abstract
% Context heading (optional), leave it empty if necessary: 
{The all-sky survey from the \textit{Planck} space telescope has revealed that thermal emission from Galactic dust is polarized on scales ranging from the whole sky down to the inner regions of molecular clouds, Polarized dust emission can  therefore be used as a probe for magnetic fields at different scales. In particular, the analysis of the relative orientation between the density structures and the magnetic field projected on the plane of the sky, can provide information on the role of magnetic fields in shaping the structure of molecular clouds where star formation takes place.} 
% Aims heading (mandatory):
{The orientation of the magnetic field with respect to the density structures has been investigated using different methods. The goal of this paper is to explicitly compare two of the methods used for this purpose: the Rolling Hough Transform (RHT) and the gradient technique.}
% Methods heading (mandatory): 
{We have applied the RHT method and the gradient technique to two specific regions, Region 1 and Region 2, identified in synthetic surface brightness maps at 353 GHz (850 $\mu$ m) generated via magneto hydrodynamic simulations post-processed using radiative transfer modelling. For both methods we have derived the relative orientation between the magnetic field and the density structures, to which we have applied two different statistics, the histogram of relative orientation (HRO) statistic and the projected Rayleigh statistic (PRS) to quantify the variations of the relative orientation as a function of column density.}
% Results heading (mandatory): 
{We find that the samples of pixels selected by each method are substantially different. When the methods are applied to the same pixel selection, the results in terms of relative orientation as a function of column density are consistent between each other, although with some noticeable differences. When each method is applied to its own pixel selection, the differences are much more apparent. Different methods applied to the same region produce opposite trends of the parameters used to quantify the behavior of the relative orientation, which in some cases are consistent with the results of previous studies. In Region 1, the RHT method roughly reproduces the observed trend of the relative orientation becoming more perpendicular for increasing column density, while the gradient method, applied at the same resolution as RHT, gives the opposite trend, with the relative orientation moving towards a more parallel alignment. In Region 2, the situation is reversed.
The inconsistent results, which are due to the different pixel selections operated by the methods and to the intrinsic differences between these latter, provide complementary valuable information.}
% Conclusions heading (optional), leave it empty if necessary:
{Our results indicate that the interpretation of dust polarization data based on the analysis of the relative orientation between magnetic field and density structures should take into account the specificity of the methods used to determine such orientation. Indeed, the combined use of complementary techniques like the RHT and the gradient methods provides a more complete information, which can be advantageously used to investigate the physical mechanisms operating in magnetized molecular clouds and responsible for the observed behaviors.}

%% Keywords should appear after the \abstract command. See the
%% instructions to authors for the journal to which you are submitting
%% your paper to determine what keyword punctuation is appropriate.


 \authorrunning{E. R. Micelotta et al.}

   \titlerunning{Dust polarization on MHD simulations}
   \maketitle

%% From the front matter, we move on to the body of the paper.
%% In the first two sections, notice the use of the natbib \citep
%% and \citet commands to identify citations.  The citations are
%% tied to the reference list via symbolic KEYs. The KEY corresponds
%% to the KEY in the \bibitem in the reference list below.

%=========================================
\section{Introduction}
%=========================================

% FIGURE  *************************************************************
%
\begin{figure*}
  \begin{center}
    \includegraphics[width=\hsize]{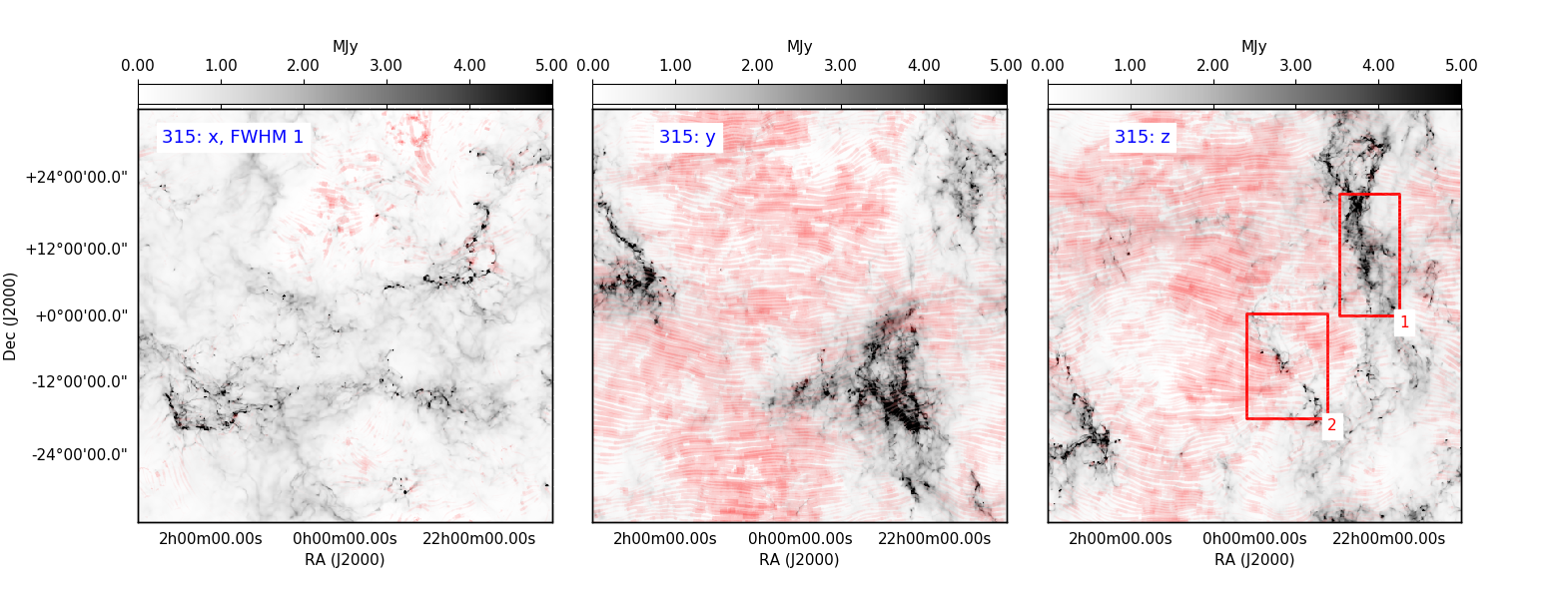} 
  \end{center}
  \caption{Surface brightness maps at 353 GHz (850 $\mu$m -- shades of
    gray) resulting
    from radiative transfer calculations performed on density structure
    maps from MHD simulations. Each frame shows the maps corresponding to
    one observing direction ($x$-left, $y$-center, $z$-right). The
    number ``315'' 
    labeling each panel 
    indicates that the snapshots have been taken 11 Myr after
    self-gravity has been switched on in the simulation. The maps have
    been convolved over a 1-pixel gaussian beam (FWHM = 1). The
    ``fingerprint'' pattern shows the direction of the magnetic field and
    the intensity of the red color increases with the polarisation
    fraction. The red boxes indicate the regions where individual analysis
    has been performed.
    }
\label{fig:polvec_combine}
\end{figure*}
% ********************************************************************

The role of magnetic fields in the formation and evolution of molecular clouds (MCs) is the object of active research, and represents one of the keys to understand the earliest phases of star formation \citep[see e.g.,][for a review]{li14}. It has been established that MCs have a filamentary nature \citep[e.g.,][]{ungerechts87, bally87, ward10, arzoumanian11}, and indeed filaments are observed to  develop in numerical simulations \citep[e.g.,][]{deAvillez05, padoan07, federrath13}. However, the exact mechanisms leading to the formation and evolution of such elongated structures, and in particular the relative contributions of turbulence, gravity and magnetic fields, are not fully understood yet.

The study of the relative orientation between the filamentary column density structures and the projection of the magnetic field in the plane of the sky can provide insights on the formation mechanisms of the filaments and ultimately, on the evolutionary stage of the star-forming region hosting them.
Different methods have been used so far to identify linear structures in astronomical data and quantify their alignment. For instance, the DisPerSE method, originally developed by \citet{sousbie11} to recover density skeletons in cosmic web data, has been used by \citet{peretto12} and \citet{palmeirim13} to study the formation of filamentary structures in the Pipe nebula and in the Taurus molecular cloud respectively. The inertia matrix has been used to investigate the physical origin of filaments in magnetized media \citep{hennebelle13}, while the Hessian matrix has been used to analyze the relative orientation between density structures and the magnetic field traced by dust polarization in the all-sky \textit{Planck} data \citep{planck_inter_XXXII_16}. 

In this paper we focus on the comparison between two other methods which have been widely used to specifically  investigate the relative orientation between column density structures and the magnetic field in which they are immersed.
The first method is the Rolling Hough Transform \citep[RHT --][]{clark14}, and the second is the gradient technique \citep{soler13, soler_Planck_XXXV_16}. Both methods recover the same quantity, i.e., the angle between the direction of the density structures and the direction of the magnetic field projected on the plane of the sky. While this latter is inferred in both cases from dust polarization, the former is derived in two different ways.    

RHT has been used for instance by \citet{koch15}, \citet{malinen16}, \citet{panopoulou16}, \citet{alina17} and in numerical simulations by \citet{inoue16}. The gradient technique has been employed to analyze a variety of observational data \citep[for instance,][]{soler_Planck_XXXV_16,soler17, jow18}  and in simulations \citep[e.g.,][]{chen16, soler_hennebelle17}. 

One of the main results which emerged from both observational and theoretical studies of the alignment between density structures and magnetic field is that the relative orientation has a bimodal distribution, with low-density column density structures preferentially aligned parallel to the magnetic field and high-density regions mostly orthogonal to the field or without a favored orientation.

In terms of this result, the RHT and the gradient techniques, which have been applied to different datasets, give results which are consistent in most of the cases. In consideration of the intrinsic differences between the two methods, this could be seen as a bit surprising. To our best knowledge, however, no systematic comparison between the two techniques has been performed so far. Such a comparison is therefore the focus of our paper, to elucidate which factors determine the differences and similarities between the output from the two methods. This is necessary to properly interpret the data and investigate the physical processes underpinning the observed interplay between gas column density structures and magnetic field.  

We perform our analysis on density structures generated via
state-of-the-art magneto hydrodynamic (MHD) simulations and traced
via their emission at 353\,GHz (850 $\mu$m). At this frequency
(wavelength) emission is due to dust grains reprocessing radiation
from an illuminating source. Because dust grains are immersed into a
magnetic field, their emission is polarized. Dust is therefore a
tracer of gas column density (through its emission) and of magnetic
fields (through its polarized emission).

This paper is organized as follows. In Sec.~\ref{sec:MHD} and Sec.~\ref{sec:RT} we describe the MHD simulation and the radiative transfer calculations, respectively. Sec.~\ref{sec:RHT_grad} outlines the methods used to quantify the orientation of the local magnetic field with respect to the density structures. Our results are illustrated in Sec.~\ref{sec:results} and discussed in Sec.~\ref{sec:disc_concl} where we also present our conclusions.

%=========================================
\section{MHD simulation}\label{sec:MHD}
%=========================================

We use an MHD simulation describing the ISM in a 250 pc region, where the turbulence is driven by supernova (SN) explosions. The simulation, carried out with the Ramses adaptive-mesh-refinement (AMR) code \citep{Teyssier02}, was first presented in \citet{Padoan+16SNI}, and further analyzed in \citet{Pan+16SNII,Padoan+16SNIII}; 
we only describe it briefly here and refer the reader to \citet{Padoan+16SNIII} 
for further details. We adopt a minimum cell size (maximum spatial resolution) 
of ${\rm d}x=0.24$ pc, periodic boundary conditions, a mean density of 5~cm$^{-3}$ 
(corresponding to a total mass of $1.9\times 10^6$ $M_{\odot}$) and a SN rate 
of 6.25~Myr$^{-1}$, with the SN explosions randomly distributed in space and time. Individual SN explosions are implemented with an instantaneous addition of 
10$^{51}$~erg of thermal energy and 15~M$_{\odot}$ of gas, distributed according 
to an exponential profile on a spherical region of radius $r_{\rm SN}=3 {\rm d}x=0.73$~pc, 
which guarantees numerical convergence of the SN remnant evolution
\citep{Kim+Ostriker15SN}. The energy equation includes the $p {\rm d}V$ work, the thermal 
energy from the SN explosions, photoelectric heating up to a critical density of 
200~cm$^{-3}$, and parametrized cooling functions from \citet{Gnedin+Hollon12}. 

The simulation starts with zero velocity, uniform density, $n_{\rm H,0}=5$ cm$^{-3}$, uniform magnetic field, and uniform temperature, $T_0=10^4$ K. The SN-driven turbulence
brings the mean thermal, magnetic and kinetic energy to an approximate steady-state, where the amplified magnetic field has an rms value of 7.2 $\mu$G and an average of $|{\rm B}|$ of 6.0 $\mu$G, consistent with the observations. The simulation was integrated for 45 Myr without self-gravity and then continued with self-gravity for 11 Myr. The snapshot used in this work is taken from the end of this second part of the simulation including self-gravity.

The analysis of this simulation has shown that SN-driven turbulence can explain the formation and evolution of molecular clouds (MCs), their internal turbulence, their lifetimes, their mass and size distribution. With a higher-resolution continuation of the simulation, including sink particles, it was shown that SN-driven turbulence may also explain the low value of the star-formation rate in MCs and the even lower global star-formation rate in the Galaxy \citet{Padoan+17SNIV}. Because of their realistic properties, MCs selected from this simulation are an ideal tool to study properties that are difficult to infer directly from the observations, such as the spatial structure of the magnetic field, and its relation to the morphology of density structures, such as filaments and clumps. 

The AMR method provides a high spatial resolution in regions of interest in the large computational volume, namely MC clouds and their substructures. However, for the purpose of the analysis of this work, we should stress the caveat that low density regions are not spatially refined in this simulations, so their spatial resolution corresponds to that of the $128^3$ root grid, that is $\sim 2$~pc. This resolution is clearly insufficient to capture the filamentary structure often observed in low-density regions surrounding MCs. Thus, the study of the relative orientation of the magnetic and thin, low-density filaments is beyond the scope of this work. However, the simulation allows us to study the orientation of the magnetic field with respect to dense structures, and to compare different methods of pursuing such a study.

%=========================================
\section{Radiative transfer calculations}\label{sec:RT}
%=========================================

We use radiative transfer (RT) modelling to predict the total dust
emission and the polarisation observed from the model clouds. The
calculations were performed with the SOC program, which is a Monte
Carlo radiative transfer program for the calculations of dust emission
and scattering. SOC has been used in some previous publications
\citep{Gordon2017, GCC-IX} and will be described in
more detail in a forthcoming paper \citep{Juvela2018}.

SOC can directly use the octtree grid of the MHD simulations. However,
in this paper the cloud data were resampled onto a regular 512$^3$
grid with a 0.49\,pc cell size. For example at a distance of 550\,pc,
this would thus correspond to a resolution of 3.35$\arcmin$. 

We adopted the dust model from \cite{Compiegne2011}. The model
cloud was illuminated from the outside by an isotropic radiation field
with the intensities given in \cite{Mathis1983}. We did not include
any discrete radiation sources inside the model volume. The average
visual extinction through the model is $A_{\rm V}$=1.7\,mag but,
because of the inhomogeneity of the density field, the large-scale
radiation field remains relatively constant throughout the volume.
Therefore, the individual dense regions within the models are
illuminated by an intensity that is only slightly weaker than the
external field. 

RT calculations were used to solve the dust temperature in each model
cell and, based on this information, to produce surface brightness
images at 353\,GHz. The maps were calculated for observers in the
directions of the main axes. In this paper, we do not model in detail
the grain alignment processes and instead assume assume a constant efficiency of dust alignment, with an intrinsic polarization fraction of 20\% 
Thus, in addition to the
total intensity $I$, we calculated maps for the Stokes parameters $Q$ and
$U$ from the following equations:
\begin{eqnarray}
Q & = & \int j_{\nu}(s) \, \cos 2\Psi(s) \,  \cos^2 \gamma(s)
\, {\rm e}^{-\tau_\nu(s)} \, R(s) \, {\rm d}s \\
U & = & \int j_{\nu}(s) \, \sin 2\Psi(s) \, \cos^2 \gamma(s) \, {\rm
  e}^{-\tau_\nu(s)} \, R(s) \, {\rm d}s,
\end{eqnarray}
where the Rayleigh polarisation reduction factor $R$ is 
constant. 
Here $j_{\nu}$ is the emissivity that depends on the local density and dust temperature.
At 353\,GHz, the optical depth between the point of emission and the observer,
$\tau_{\nu}(s)$, is very small and the exponential terms are
practically equal to 1. The
equations include two angles that depend on the direction of the magnetic field $\mathbf
B$: $\Psi$ is the position angle of $\mathbf B$ projected onto the plane of the sky
(with respect to a chosen reference direction) and $\gamma$ is the angle between
$\mathbf B$ and the plane of the sky.

Given maps of $Q$ and $U$, one can further estimate the polarisation angle $\chi$ from
\begin{equation}
\tan 2 \chi = U/Q,
\end{equation}
and the polarisation fraction $p$ from
\begin{equation}
p = \sqrt{Q^2+U^2}/I.
\end{equation}
We do not include any additional corrections for noise-induced bias in
these quantities \citep[see][]{Montier2015}. The noise of the
synthetic $I$, $Q$, and $U$ maps is small. Furthermore, because the
Monte Carlo errors are contained in the emissivity $j_{\nu}$, these
leave both $\chi$ and $p$ essentially unchanged, apart from small
variations in the relative weighting of different cells along the line of
sight.

The calculated surface brightness maps at 353\,GHz are shown in
Fig.~\ref{fig:polvec_combine}. Each panel refers to one observing direction ($x$-left,
$y$-center, $z$-right). We consider specifically the snapshots
labelled as ``315'', indicating that they have been taken 11 Myr after
self-gravity has been switched on in the simulation. This ensure that
the initial density field had time to collapse into denser
structures. The resolution of the maps is given by 
convolution over a 1-pixel Gaussian beam (FWHM = 1). The surface
brightness maps are overlaid with a red ``fingerprint'' pattern. The
pattern itself indicates the direction of the magnetic field, while
the intensity of the red color increases with the polarisation
fraction.

The lack of polarisation along the $x$-axis reflects the initial
orientation of the magnetic field along this direction. In the $y$ and
$z$ directions, the magnetic field shows a clear large-scale
orientation in the more diffuse regions, while the polarisation
fraction is strongly reduced in the denser structures. To perform our
analysis, we have selected two regions whose physical size is
comparable to those of the molecular clouds analysed in
\citet{soler_Planck_XXXV_16}. Our selected regions are highlighted in
Fig.~\ref{fig:polvec_combine}: Region 1 is almost entirely occupied by
a dense and thick filamentary structure with some branches, while
Region 2 includes a thinner filament with denser knots surrounded by a
more tenuous medium.

\section{How to quantify the relative orientation between magnetic
field and density structures}\label{sec:RHT_grad}

\subsection{The Rolling Hough Transform method}

The Rolling Hough Transform (RHT) has been introduced by
\citet{clark14} and it is based on the Hough transform from
\citet{hough62} as implemented by \citet{duda72}. The Hough transform
was originally devised to track the motion of particles in high-energy
physics experiments. \citet{clark14} developed a rolling version of
the Hough transform and used it to identify filamentary structures in
H{\sc I} data. The fundamental idea is the following. The image to
analyze is high-pass filtered with a filter $D_K$ to enhance the
filamentary structure, and the image is then thresholded to form a
bitmask. A patch of size $D_W$ is centered on each pixel and rolls
across the data. The number of
pixels inside the extraction window that pass the threshold and that are aligned with a local
direction $\theta$ is identified. Finally the direction $\theta_\text{RHT}$ for
which the maximum number of pixels pass a minimum threshold $Z$
is elected as being the local filament direction \citep[Fig.~2
in][]{clark14}. With this we can derive directly the angle
$\phi_\text{RHT}$ between the direction of the magnetic field and the filament:
\begin{equation}
	\phi_\text{RHT} = \theta_\text{RHT} - \Psi\,,
\end{equation}
where $\Psi$ is the position angle of $\mathbf B$.

\subsection{The gradient technique}

The gradient method consists in finding the purely local orientation of the
magnetic field with respect to the 
column density isocontour. This is equivalent to calculating the local
orientation of the unit polarisation pseudo-vector $\vec{\hat{P}}$ (orthogonal to the
magnetic field) with respect to the local gradient, which is by
definition perpendicular to the column density isocontour. In
practice, the gradient derives directly from the finite
differenciation of the intensity field on a mesh.

The unnormalized gradient of the column density $N_{\rm H}$ is given by  $N_{\rm H}$
\begin{equation}
  \vec{G}(\vec{a}) = \nabla_{\vec{x}} N_{\rm H} \equiv \frac{1}{2} \left(\begin{array}{c}
          N_{\rm H\, i+1,\,j} - N_{\rm H\, i-1,\,j} \\
          N_{\rm H\, i,\,j+1} - N_{\rm H\, i,\,j-1}
  \end{array}\right)\,,
\end{equation}
with $\vec{a}=(i,j)$.
The gradient is unnormalized as there should be a scaling factor to
account for grid mesh spacing in the finite differenciation. However,
such a scaling is unimportant as we are only interested in the unit
vector giving the direction
\begin{equation}
  \vec{\hat{G}} = \frac{\vec{G}}{||G(\vec{a})||}\,.
\end{equation}
The polarization angle $\chi$ (Sec.~\ref{sec:RT}) characterises the
direction of $\vec{\hat{P}}$ and is defined as
\begin{equation}
  \chi(\vec{a}) = \frac{1}{2} (\text{arctan2}(Q(\vec{a}), U(\vec{a})) + \pi),
\end{equation}
from which we derive the unit pseudo-vector $\vec{\hat{P}}$:
\begin{equation}
  \vec{\hat{P}}(\vec{a}) = \left(\begin{array}{c} \cos(\chi(\vec{a})) \\ \sin(\chi(\vec{a}))
\end{array}\right)
\end{equation}
From the above equations, we obtain the relative angle between the
normal to the column density and the polarisation direction \citep[see
Eq.~2 in][]{soler_Planck_XXXV_16}:
\begin{equation}\label{eq:phi_grad}
  \phi_\text{grad}(\vec{a}) = \text{arctan}\left( \vec{\hat{G}}(\vec{a}) \times \vec{\hat{P}}(\vec{x}), \vec{\hat{G}}(\vec{a}) \cdot \vec{\hat{P}}(\vec{a}) \right)\,,
\end{equation}
which is equivalent to the angle between the direction of the magnetic
field and the tangent to the column density isocontour.  From
Eq.~\ref{eq:phi_grad} it is clear that the absolute amplitudes of the
two vectors do not matter as only the value of their ratio is really
used.

% Region selection
The pixels over which we perform our analysis are selected applying
the following basic criterion on the above mentioned set of
coordinates $\mathcal{S}$:
\begin{equation} \mathcal{S}_\text{grad} = \mathcal{S}_\text{region}
\cap \{ \vec{a}\, |\, ||\vec{G}(\vec{a})|| > G_\text{threshold} \},
\end{equation} 
with $\mathcal{S}_\text{region}$ being the set of
pixels included in our selected Region 1 and Region 2 (Fig.~\ref{fig:polvec_combine}).  We define the threshold for
the gradient as
\begin{equation} G_\text{threshold} =
\frac{1}{|\mathcal{S}_\text{ref}|} \sum_{\vec{a}
\in\mathcal{S}_\text{ref}} ||\vec{G}(\vec{a})||,
\end{equation} CL
with $\mathcal{S}_\text{ref}$ being the 
reference region, for which we chose a smooth and low-brightness area in the y-axis snapshot, centered on RA=0h00m00s; Dec= +12$^\circ$00'00" (J2000) and with a similar size than Regions 1 \& 2.

\subsection{Histograms of relative orientations and the shape parameter $\xi$}

We use the calculated values of $\phi_\text{RHT}(\vec{a})$ and
$\phi_\text{grad}(\vec{a})$ to build the histograms of relative
orientations \citep[HROs -- e.g.,][]{soler13, soler_Planck_XXXV_16,
planck_inter_XXXVIII_16, malinen16} for each of the 15 bins in which
we grouped our column density values. The bins have been designed to contain the same number of pixels. The HROs are here presented (Sec.~\ref{sec:rel_orient}) as a
probability density $P(\phi)$ versus $\phi$, where $P(\phi)$ is
obtained via normalisation of $\tilde{A}_i$:
\begin{equation}
	P(\phi) = \frac{\tilde{A}_i}{h_{i+1} - h_i} \text{for }h_{i}
        \le \phi < h_{i+1}\,.
\end{equation}
A histogram peaking at around $\phi$ = 0$^\circ$ means that the
magnetic field is preferentially oriented parallel to the density
structures, while a histogram having a minimum at the same angle
indicates that the magnetic field is mostly orthogonal to the density
structures.  

From the HROs we compute the
histogram shape parameter $\xi$ \citep{soler_Planck_XXXV_16, soler17} to
quantify the variations occurring in the HROs as a function of column
density. The problem can be formalised in a general way as follows. 
% HRO statistics
Based on a set $\mathcal{S}$ of $\vec{a}$ coordinates, we calculate the
statistics of mean values of $\phi_\text{RHT}$ and
$\phi_\text{grad}$. For each of the two angles we compute:
\begin{equation}
	\tilde{A}_i = \left| \{ \vec{a} \in \mathcal{S}\, |\, h_i \le \phi(\vec{a}) < h_{i+1} \} \right|,
\end{equation}
with $h_i = -90 + 180 \times i / 12$.
We further define $A_0$ and $A_{90}$ as in \citet{soler17}:
\begin{align}
  A_0 & = \tilde{A}_0 + \tilde{A}_1 + \tilde{A}_{11} + \tilde{A}_{12} \\
  A_{90} & = \tilde{A}_4 + \tilde{A}_6 + \tilde{A}_7 +\tilde{A}_8
\end{align}
Finally, we define $\xi$, as in \citet{soler17}:
\begin{equation}
   \xi = \frac{A_0 - A_{90}}{A_0 + A_{90}}
\end{equation}
To get an approximation of the error bar due to Poisson count we use
the Gaussian limit of the Poisson law. For a Poisson intensity
$\lambda$ the standard deviation is $\sqrt{\lambda}$. The Gaussian
error on $\xi$ is
\begin{equation}
   \sigma_\xi = \sqrt{\frac{4 A_0 A_{90}}{(A_0 + A_{90})^3}}.
\end{equation}
For each column density bin, $\xi > 0$ means that the magnetic field
and the density structures are mostly parallel (concave histogram),
$\xi < 0$ indicates a mostly perpendicular orientation (convex
histogram) and $\xi \approx 0$ reveals that there is no preferred
orientation (flat histogram).

\subsection{The projected Rayleigh statistics}

Besides the HRO statistics described above, we also applied to our
simulated data the so-called projected Rayleigh statistic \citep[PRS
-- ][]{jow18}. The PRS is based on the classic Rayleigh test typically
used in circular statistics to determine whether the angles of a
specific set are uniformly distributed \citep[e.g.,][]{batschelet81, glimm96,
  mardia99} and can be considered a specific case of the V statistics
\citep{durand58, mardia99}. The V statistics allows to test for
uniformity against a specific mean direction characterised by an angle
$\theta$. The V statistics for the case $\theta = 0$, corresponding to
a parallel orientation, has been renamed PRS. Following \citet{jow18}
we use our angles $\phi(\vec{a})$ (from both the RHT and gradient
technique) to define the PRS for the same set $\mathcal{S}$. The
parameter used to quantify the relative orientations is $Z_{\rm Jow}$
defined as
\begin{equation}
   Z_{\rm Jow} = \sqrt{\frac{2}{|\mathcal{S}|}} \sum_{\vec{a} \in \mathcal{S}} \cos(2\phi(\vec{a}))
\end{equation}
As from \citet{jow18}, the typical ``error'' on this quantity is
given by:
\begin{equation}
   \sigma_Z = \sqrt{\frac{2}{|\mathcal{S}|} \sum_{\vec{a} \in \mathcal{S}} \cos^2(2\phi(\vec{a}))}.
\end{equation}
As for $\xi$, positive values of $Z_{\rm Jow}$ indicate a parallel
orientation, negative values a perpendicular orientation and values
compatible with zero show that there is no preferred orientation. 

% FIGURE  *************************************************************
%
\begin{figure*}
  \begin{center}
\includegraphics[width=\hsize]{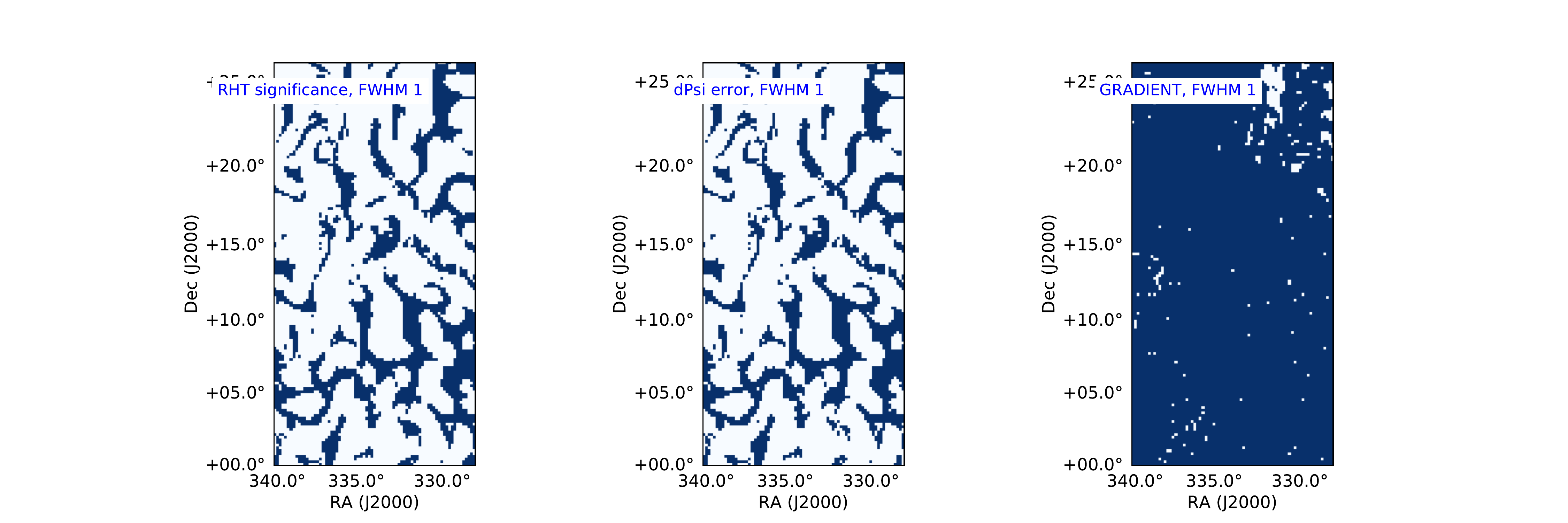}
\includegraphics[width=\hsize]{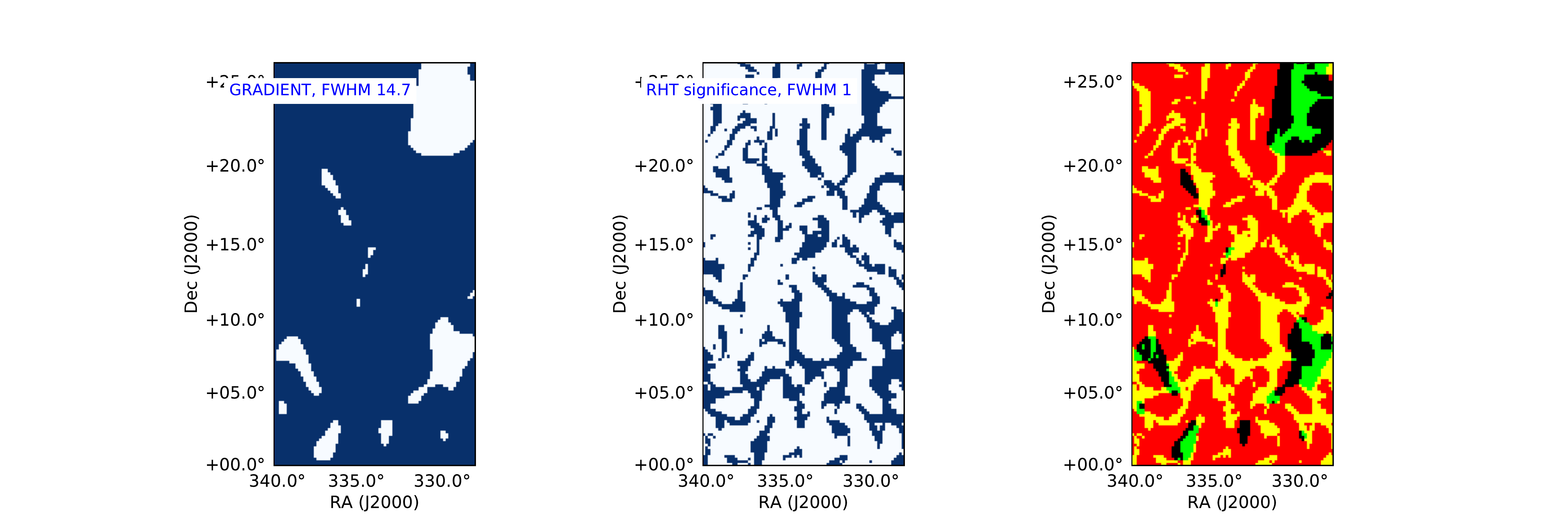}
  \end{center}
  \caption{Comparison between pixel selections in Region 1, obtained
    from the RHT and gradient methods with different criteria. The
    selected pixels are indicated in blue. Top row,
    left panel: pixels from the RHT method with significance
    of the position angle greater than 0.9; middle panel:
    pixels from RHT imposing the additional criterium that the error of
    the polarisation angle must be lower than 1$^\circ$. 
    In both cases,
    FWHM=1; right panel: selected pixel from the 
    gradient method imposing that the module of the gradient must be
    greater than the average over a reference (smooth) region, but
    keeping FWHM=1. Bottom row, left panel: selected pixel from 
    the gradient method with the same condition as above, but
    smoothing the intensity first to FWHM=14.7;
    middle panel: RHT pixel
    selected using the significance only (same as top-left) and (right
    panel) overlap
    between the first two bottom-row panels: red corresponds to the
    gradient-selected pixels (with FWHM=14.7), green to the
    RHT-selected pixels (significance only), yellow shows the
    overlapping pixels from the two methods and black the pixels
    discarded by both methods. See text for detailed explanations.}
    \label{fig:RHT_vs_GRAD_1} 
\end{figure*}
% ********************************************************************

% FIGURE  *************************************************************
%
\begin{figure*}
  \begin{center}
\includegraphics[width=\hsize]{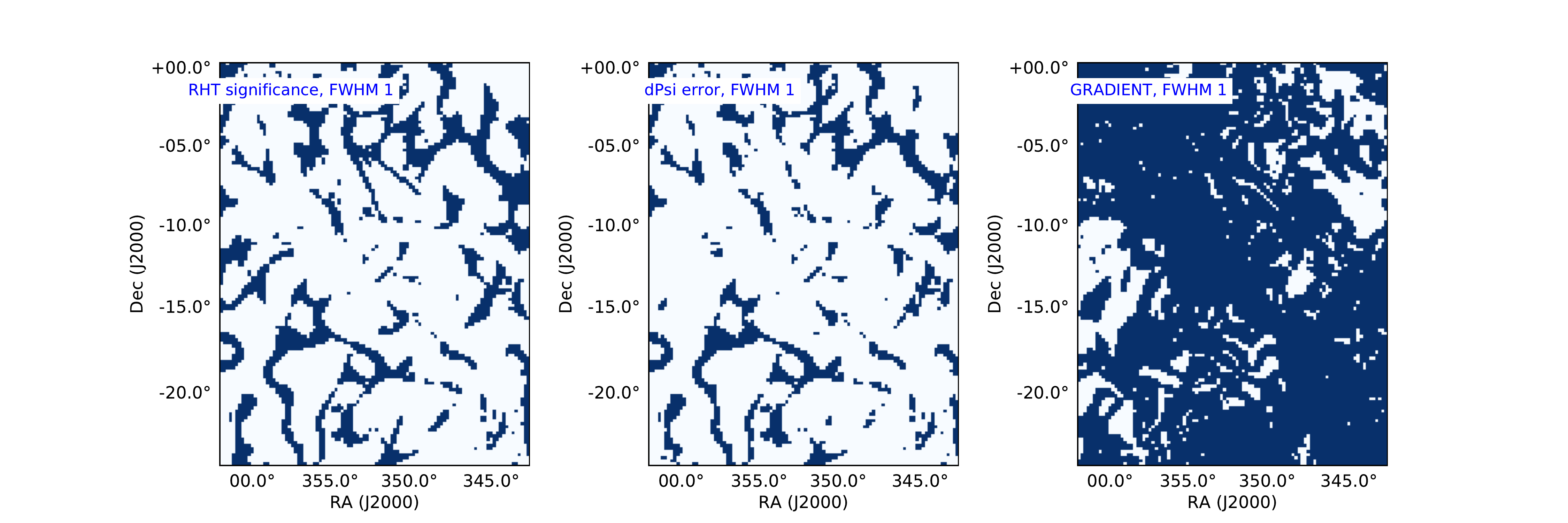}
\includegraphics[width=\hsize]{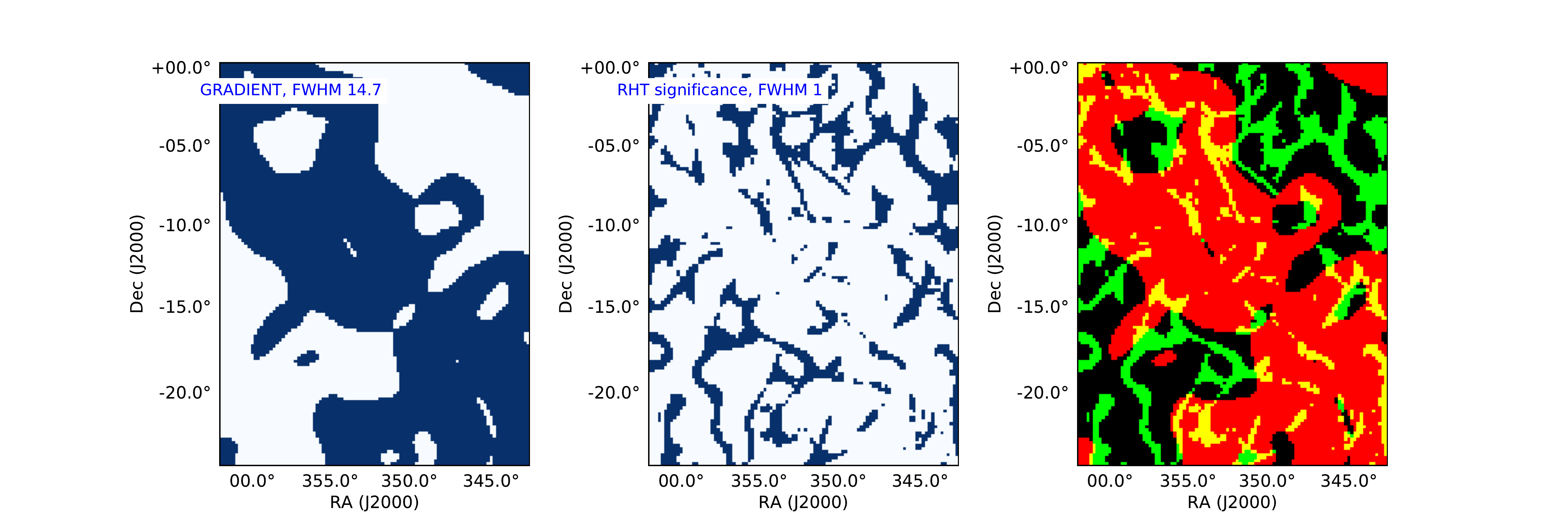}
  \end{center}
  \caption{Same as Fig.~\ref{fig:RHT_vs_GRAD_1} but for Region 2.}
    \label{fig:RHT_vs_GRAD_2} 
\end{figure*}
% ********************************************************************

% FIGURE  *************************************************************
%
\begin{figure*}
  \begin{center}
    \includegraphics[width=\hsize]{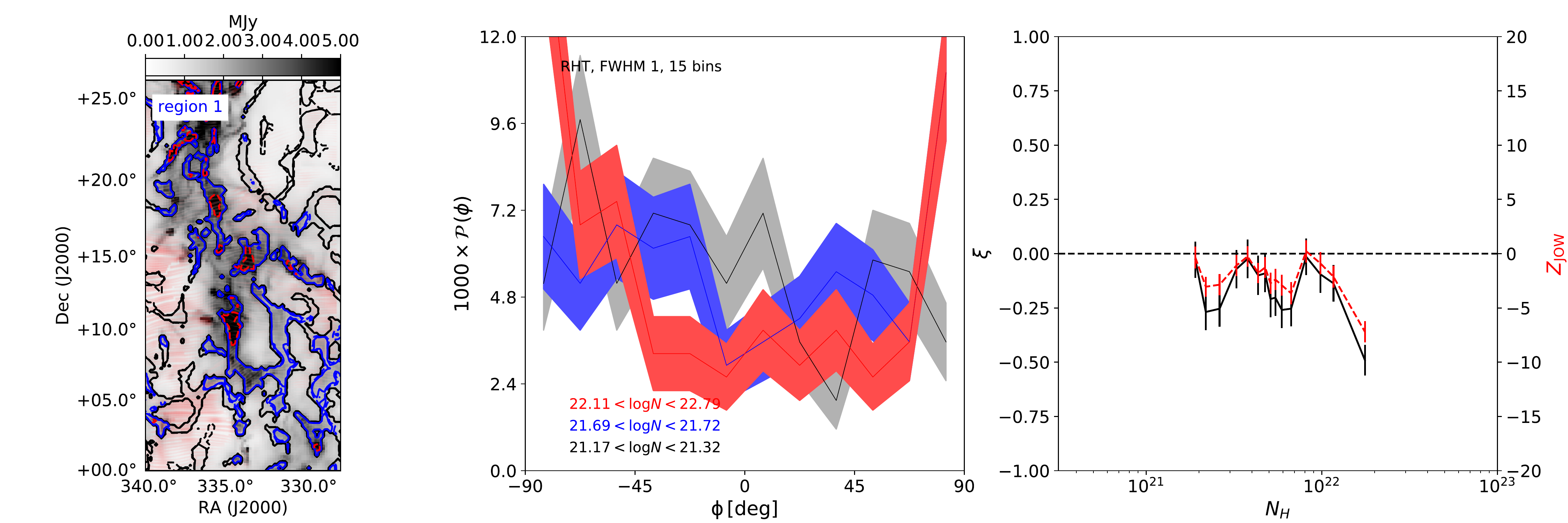}
    \\
    \includegraphics[width=\hsize]{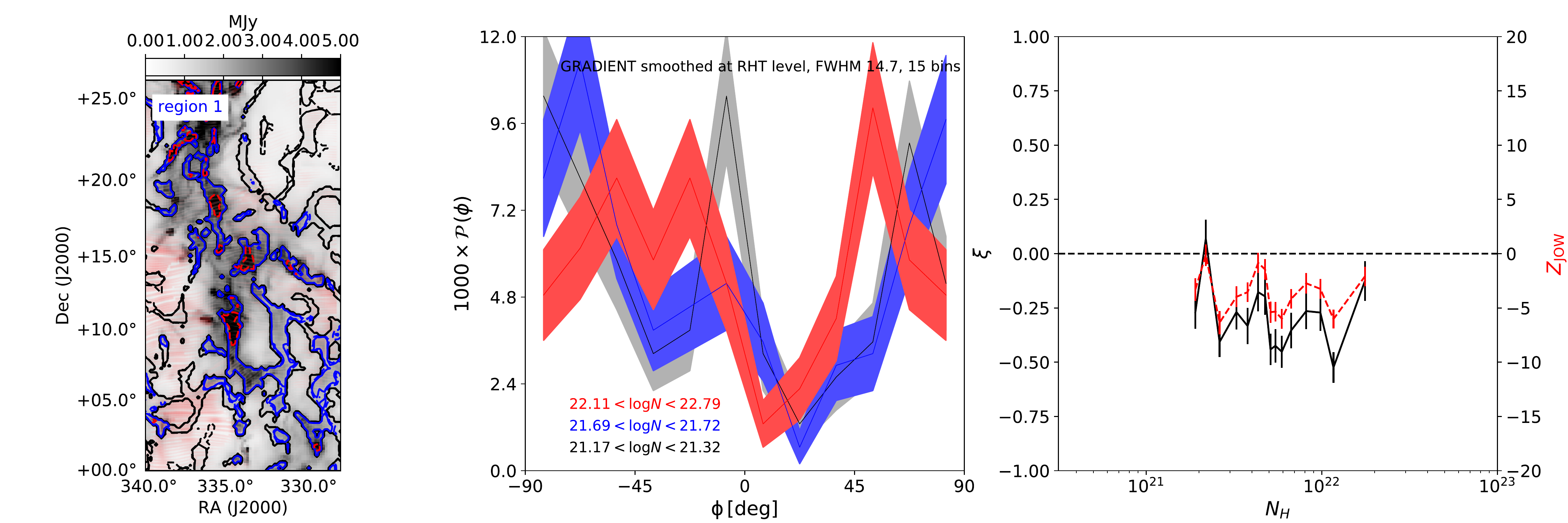} \\
    \includegraphics[width=\hsize]{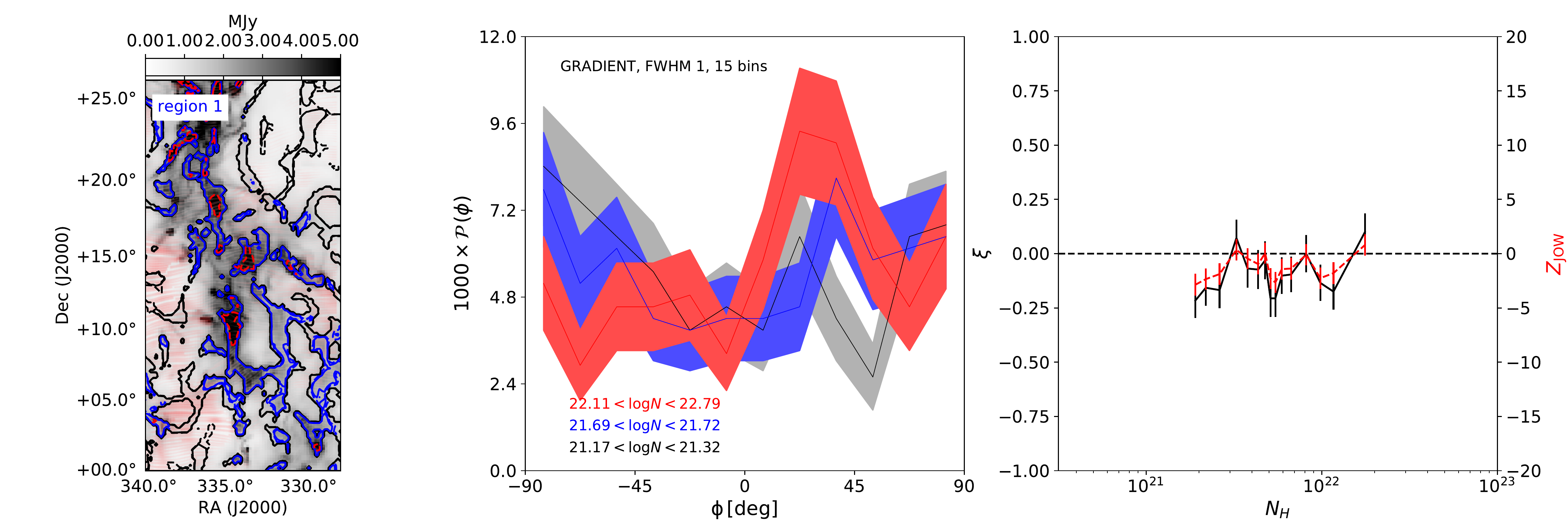}
    \\
  \end{center}
  \caption{Analysis of Region 1 from
    Fig.~\ref{fig:polvec_combine}: surface
    brightness map overlaid with the red fingerprint pattern (left
    column); histograms of relative orientation (HROs -- middle
    column) between the magnetic field direction and the density
    structures; $\xi$ and $Z_{\rm Jow}$ parameters plotted as a
    function of column density, $N_{\rm H}$ (right column). For both
    parameters, a positive value indicates a preferred parallel
    orientation, a negative value a preferred perpendicular
    orientation and a value compatible with zero indicates that there
    is no preferred orientation.
    The regions between the
    solid and dashed contours on the maps (or inside them, if only one
    contour is present) highlight the pixels corresponding to the
    column density bins shown in the HRO plots. Red and black identify 
    the highest and lowest bins respectively, while blue indicates the
    intermediate bin between the two above.
    In the HRO plots, a $\pm 1 \, \sigma$ error is
    represented by the width of the shaded areas. The 
    RHT and gradient methods have been applied to the same
    subsample of pixels selected by RHT adopting the significance
    criterium only. The
    top row shows the results from the RHT technique, the middle row
    those from the gradient technique with resolution
    matching the one of RHT (FWHM=14.7) and the bottom row those 
    from the gradient method but with FWHM=1.}
    \label{fig:same_pix_reg1} 
\end{figure*}
% ********************************************************************

\section{Results}\label{sec:results}

\subsection{Pixel selections}\label{sec:pix_sel}

The pixels selected using the RHT and gradient techniques are shown in
Fig.~\ref{fig:RHT_vs_GRAD_1} and Fig.~\ref{fig:RHT_vs_GRAD_2} for
Region 1 and Region 2 respectively. 

Among the pixels identified by the RHT method we have selected only
those with significance of the position angle greater than 0.9, as
shown in the left panel of the top row. In the middle panel, we have imposed the
additional criterium that the error of the
polarisation angle must be lower than some threshold, for which we
have adopted a reasonable value of 
1$^\circ$.
In both cases, the resolution of the brightness maps has been kept at 
FWHM=1. The differences between the two panels are minimal in both
regions. The right panel shows the pixels selected by the
gradient method and imposing
that the module of the gradient must be greater than the average over
a reference smooth region (Sec.~\ref{sec:RHT_grad}). In this case as well, the resolution of the
map is such that FWHM=1. With respect to the RHT method, the number of
selected pixels is much higher. 

The left
panel of the bottom row shows the pixels selected by the gradient
method imposing the same condition
as above, in this case, however, we have first smoothed the intensity 
to FWHM=14.7. This is to ensure that the gradient is calculated over a region
which matches the size of the ``extraction'' region in the RHT method,
where this region has diameter $D_W$. Indeed, making sure that the two
methods are working at the same resolution, i.e., on the same scale,
is necessary to perform a meaningful comparison.    

To derive the value FWHM=14.7 we used the following relations:
\begin{equation}
  {\rm FWHM} = 2\sqrt{2\ln 2} \, \sigma \approx 2.355\, \sigma,
\end{equation}
where $\sigma$ is the standard deviation of the Gaussian
distribution. We made the reasonable assumption that the
``extraction'' performed in the RHT method resembles a top-hat
filtering \citep[compare with Fig.~2 in][]{clark14}. To have Gaussian
and top-hat behaving in the same way at large scales, we need to have
$\sigma = R/2$ where $R$ is the radius of the top-hat filter. In our
case, we have $R = D_W/2$. Therefore:
\begin{equation}
 {\rm FWHM} \approx 2.355 \times \frac{D_W}{4} = 0.588\,  D_W.
\end{equation} 
Adopting $D_W =25$ (which corresponds to $D_K =10$) we get
FWHM=14.7. This should ensure that we are applying the two methods
using the same resolution. 

The comparison between the smoothed gradient and RHT selections
(bottom row, right panel) reveals that the pixels picked by the first
method are much more numerous and they occupy contiguous locations,
while the fewer RHT-selected pixels are arranged in more disperse
filamentary structures. In terms of the location of the selected
pixels, the two samples show some degree of
complementarity, which is more evident in Region 2
(Fig.~\ref{fig:RHT_vs_GRAD_2}). The gradient selection clearly follows
the denser filament which goes across the top-left/bottom-right
diagonal of the region (Fig.~\ref{fig:polvec_combine}), while the RHT
selection seems to avoid this same diagonal.

\subsection{Relative orientations}\label{sec:rel_orient}

\begin{figure*}
  \begin{center}
    \includegraphics[width=\hsize]{polvec_RHT_region_1_FWHM_1_DK_10_POL_0} \\
    \includegraphics[width=\hsize]{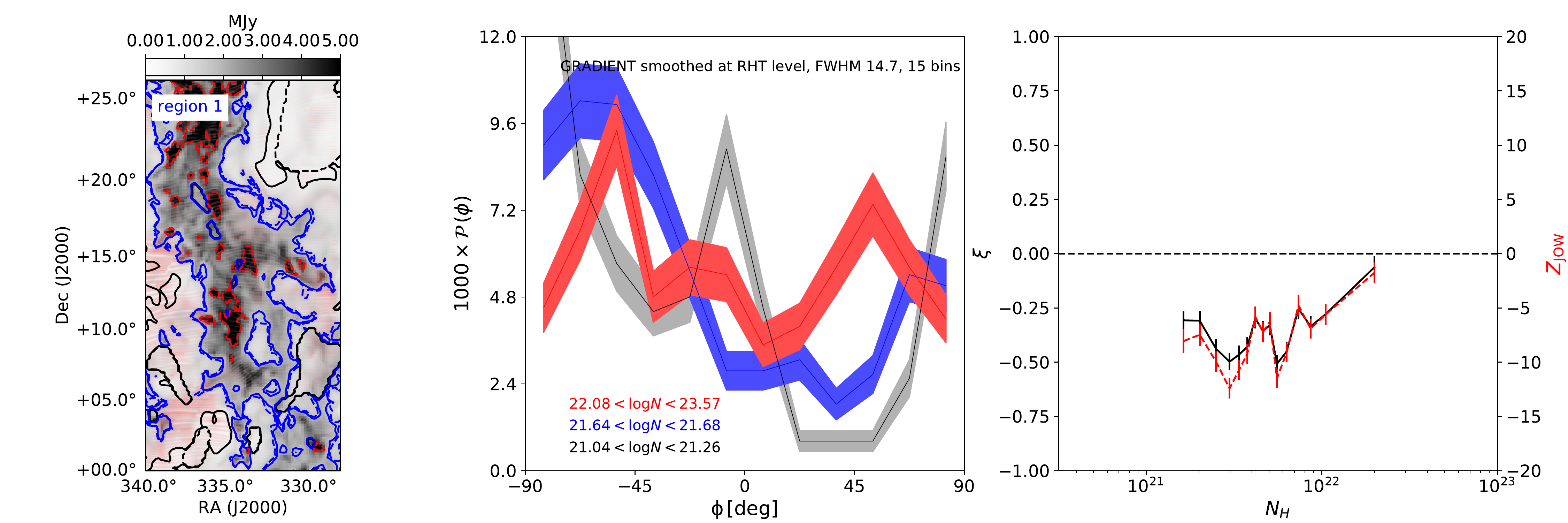} \\
    \includegraphics[width=\hsize]{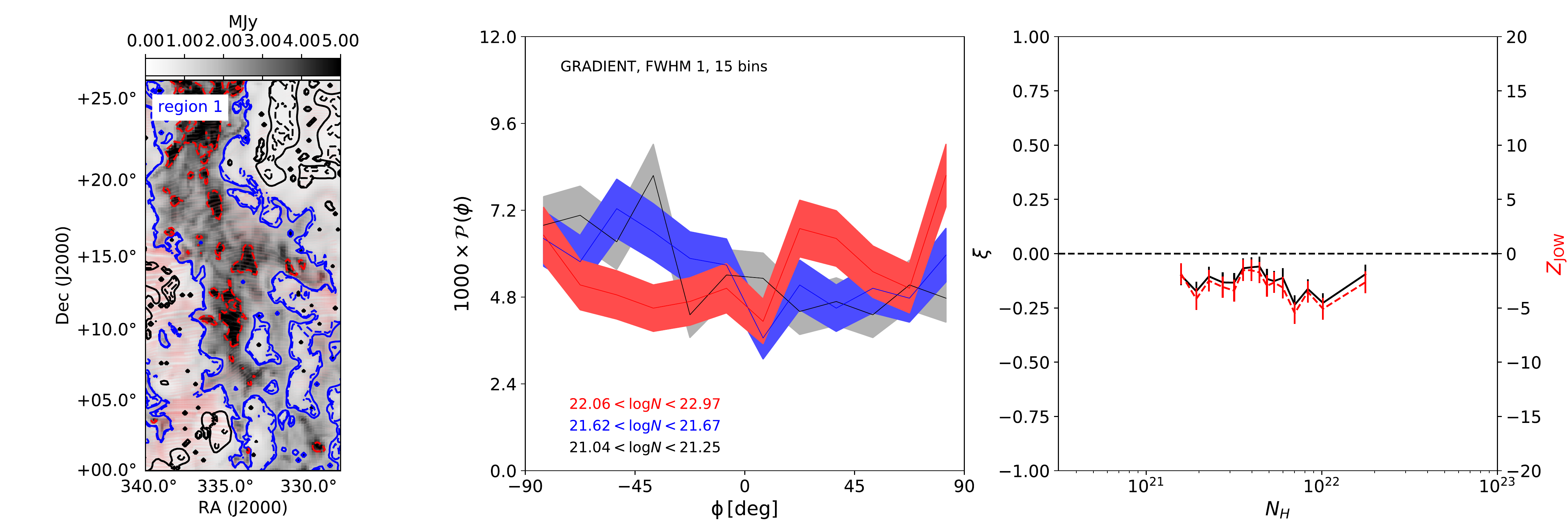} \\
  \end{center}
  \caption{Same as 
    Fig.~\ref{fig:same_pix_reg1} but in this case the RHT and gradient
    techniques have been applied each to their own native pixels
    selections, as shown by the three different panels in the left column.
    In the top row, the
    RHT technique has been applied to RHT-selected pixels
    (significance criterium).
    In the middle row, the pixels
    have been selected using the gradient method (module of the
    gradient greater than the average over a reference region) and
    smoothing the resolution to match the one of RHT (FWHM=14.7). In
    the bottom row, the gradient selection and techniques have been
    applied keeping FWHM=1.}
    \label{fig:diff_pix_reg1} 
\end{figure*}

\begin{figure*}
  \begin{center}
    \includegraphics[width=\hsize]{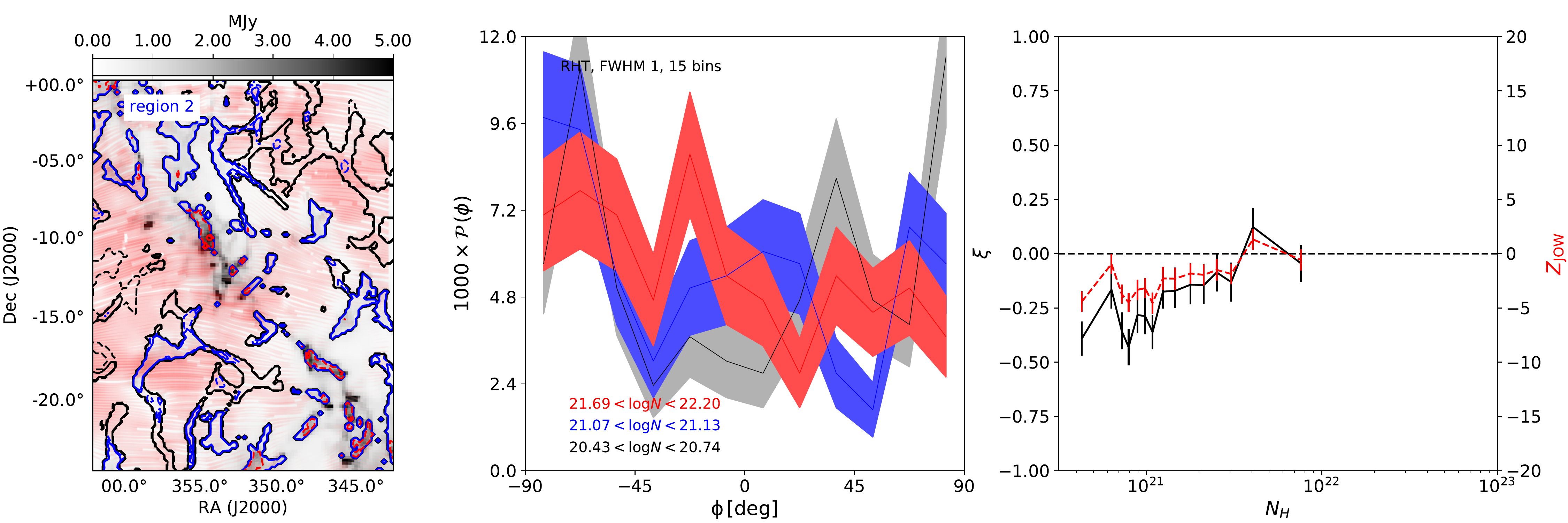}
    \\
    \includegraphics[width=\hsize]{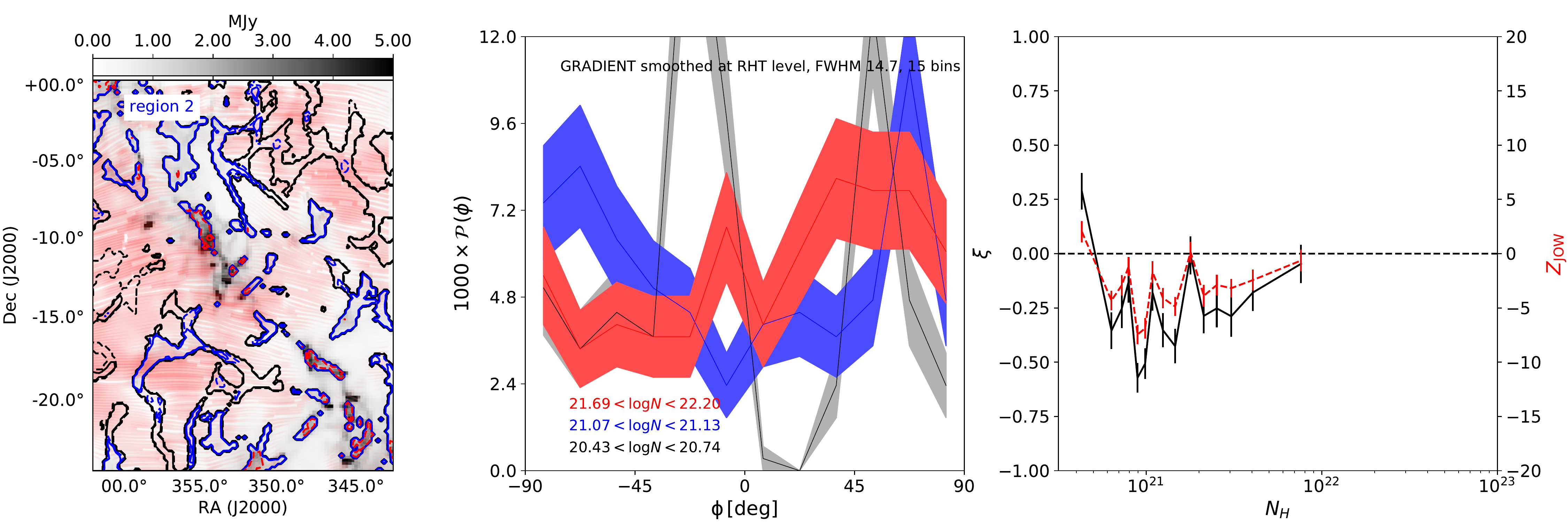} \\
    \includegraphics[width=\hsize]{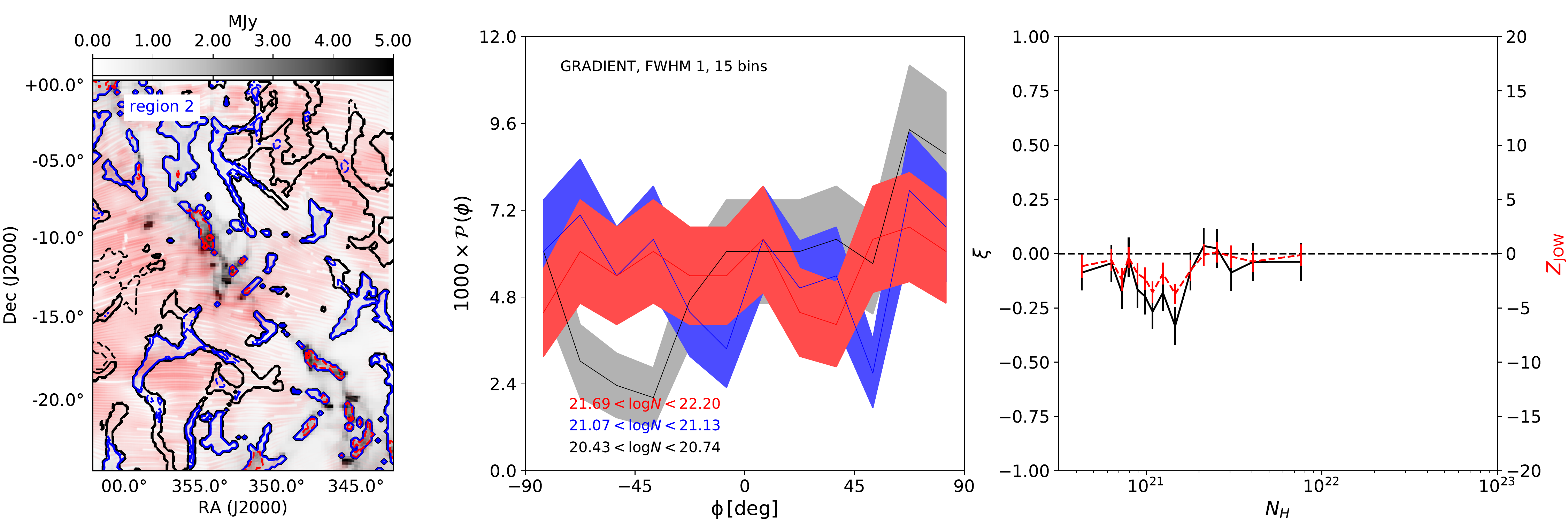}
    \\
  \end{center}
  \caption{Same as Fig.~\ref{fig:same_pix_reg1} but for Region 2.}
    \label{fig:same_pix_reg2} 
\end{figure*}

\begin{figure*}
  \begin{center}
    \includegraphics[width=\hsize]{polvec_RHT_region_2_FWHM_1_DK_10_POL_0} \\
    \includegraphics[width=\hsize]{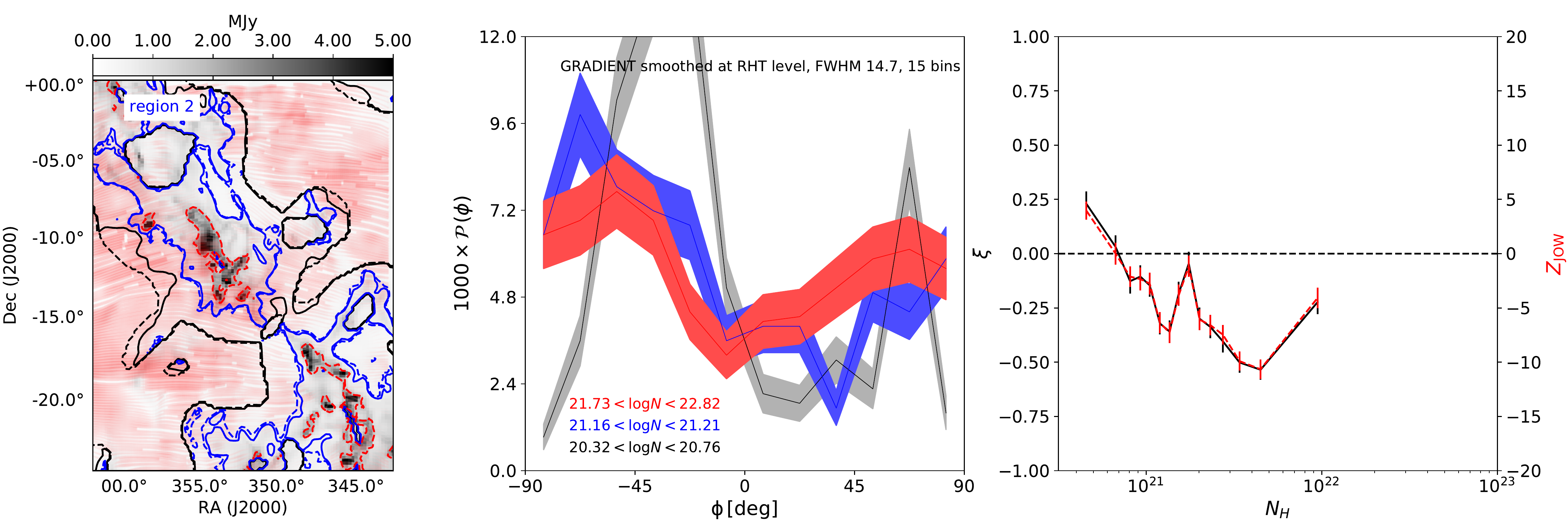} \\
    \includegraphics[width=\hsize]{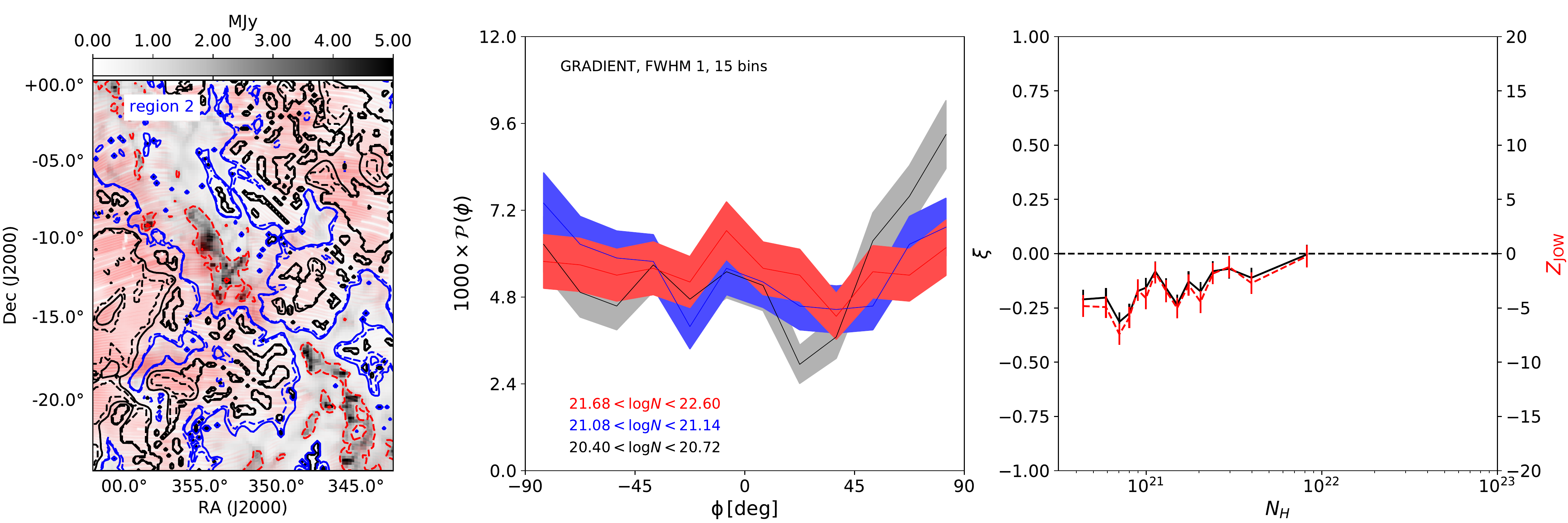} \\
  \end{center}
  \caption{Same as 
    Fig.~\ref{fig:diff_pix_reg1} but for Region 2.}
    \label{fig:diff_pix_reg2} 
\end{figure*}

Figures~\ref{fig:same_pix_reg1} to \ref{fig:diff_pix_reg2} present our
results in terms of quantification of the relative orientation between
the magnetic field and the column density structures in our simulated
regions. 

In each figure, the left column shows the intensity map of the region
overlaid with the red drapery pattern and with a set of solid and
dashed contours. The areas between the solid and dashed contours (or
inside them, if only one contour is present) identify three specific
column density bins: red is the highest one, black the lowest one and
blue the intermediate one. The central column shows the HROs for each
of the column density bins described above. The width of the shaded areas represents a $\pm 1 \, \sigma$
error. Finally, the right column presents the parameters $\xi$ and
$Z_{\rm Jow}$ defined in Sec.~\ref{sec:RHT_grad}, plotted as a function of the column
density, $N_{\rm H}$, which has been grouped into 15 bins.

We have recovered the relative orientations applying the RHT and
gradient techniques to the different pixel
selections described in Sec.~\ref{sec:pix_sel}. Specifically, in
Fig.~\ref{fig:same_pix_reg1} (for Region 1) and in
Fig.~\ref{fig:same_pix_reg2} (for Region 2) we have applied the
different techniques to the same sample, in order to evaluate the
methods having for all of them the same bias on the selected
pixels. The three panels in the first column are equal, to
illustrate that we are considering the same sample of pixels. For our
test sample, we have adopted the selection made by the RHT method with
the only additional criterium that the significance of the position
angle must be greater than 0.9. 
In the
top row, the relative orientations, both in terms of histograms and
values of the parameters $\xi$ and $Z_{\rm Jow}$, have been calculated
using the RHT method. In the middle row, we have applied the
gradient method, but first we have smoothed the resolution of the
selected pixels to FWHM=14.7, to ensure that both methods are working
on the same scale (see Sec.~\ref{sec:pix_sel}). In the bottom row, we
used the gradient method but keeping the resolution of the sample at
the original value of FWHM=1. 

In Region 1, the global preferred orientation for the density structures is orthogonal to the magnetic field, as can be deduced from the HROs and the $\xi$ and $Z_{\rm Jow}$ parameters plots. In these latter, the values of the two parameters remain negative or close to zero for most of the column density bins, revealing the perpendicular privileged orientation. The $\xi$ and $Z_{\rm Jow}$ curves are similar for the three considered cases and also between each other, showing a flat behaviour with some oscillations. This similarity is consistent with the fact that the analysis is performed on the same pixels sample. However, the errors on $Z_{\rm Jow}$ are smaller than those on $\xi$.  When the gradient method is applied after smoothing (middle row), the oscillations are amplified and the values of $\xi$ and $Z_{\rm Jow}$ are more negative than in case when the FWHM is kept at 1 (bottom row). The behavior of the highest column density bin changes depending on the method used. For RHT, the values of both parameters decrease sharply, continuing the trend of the neighbors, while for the gradient technique, they reverse the trend going toward a less negative or even positive value.     

In Region 2 (Fig.~\ref{fig:same_pix_reg2}) the preferred orientation of the density structures is again perpendicular to the direction of the magnetic field. The global trend for $\xi$ and $Z_{\rm Jow}$ as a function of $N_{\rm H}$ is flat when the gradient method is applied to data with FWHM=1 (bottom row). When the gradient method is applied to smoothed data (FWHM=14.7) the lowest column density pixel has positive $\xi$ and $Z_{\rm Jow}$, but the others have negative values which tend to increase for increasing column density. The same rising trend is observed for the RHT technique. This implies that the preferred orientation is moving towards being less perpendicular in denser regions, which is in contrast with current findings \citep[e.g.,][]{soler_Planck_XXXV_16}.  

Figures~\ref{fig:diff_pix_reg1} \& \ref{fig:diff_pix_reg2} (for Region
1 and Region 2 respectively) are the analogues of
Figs.~\ref{fig:same_pix_reg1} \& \ref{fig:same_pix_reg2} but the
relative orientations have been derived applying each method to its own native
pixel selection, as illustrated by the three different panels in the
left column. Nothing changes for the top row, where we used the RHT
method on the RHT-selected pixels. In the middle row, the pixels have
been selected by the
gradient method and imposing a threshold on the module of the
gradient. Then, we lowered the resolution of
the image to FWHM=14.7 and we applied the gradient technique to derive
the corresponding HROs and $\xi$ -- $Z_{\rm Jow}$ plots. Finally, 
the bottom row shows the results of using the gradient technique over
gradient-selected pixels, but keeping the resolution of the selection
at FWHM=1. 

In Region 1, the preferred orientation of the density structures is again perpendicular to the magnetic field. The gradient method applied to its native selection with no smoothing (bottom row) gives for $\xi$ and $Z_{\rm Jow}$ the same flat behavior as a function of the column density recovered in the previous case, but with the amplitude of the oscillations and the error bars strongly reduced. On the other end, the gradient method applied to the smoothed pixels sample (middle row) show a well defined increasing behavior, given in particular by the highest column density bin, in contrast with the trend generally observed. 

In Region 2, the full resolution gradient selection (FWHM=1) provides $\xi$ and $Z_{\rm Jow}$ which this time tend to rise towards less negative values for increasing $N_{\rm H}$, similarly to the RHT case but with reduced errors. When the gradient method is applied to the smoothed data (middle row) the behaviors of $\xi$ and $Z_{\rm Jow}$ show a remarkable difference with respect to the previous cases. The two curves, which are essentially coincident, start at a positive value for the lowest column density bin and show a clear decrease towards the negative region of the plot, except for a peak just before 2$\times$10$^{21}$ cm$^{-2}$ and for the highest column density bin which reverts the trend moving towards less negative values. With the exception of the bump and the last bin then, we recover in this case the expected transition of the preferred orientation of the density structures, going from mostly parallel to the magnetic field in tenuous media to mostly perpendicular in denser regions.   

\begin{figure*}
  \begin{center}
    \includegraphics[width=\hsize]{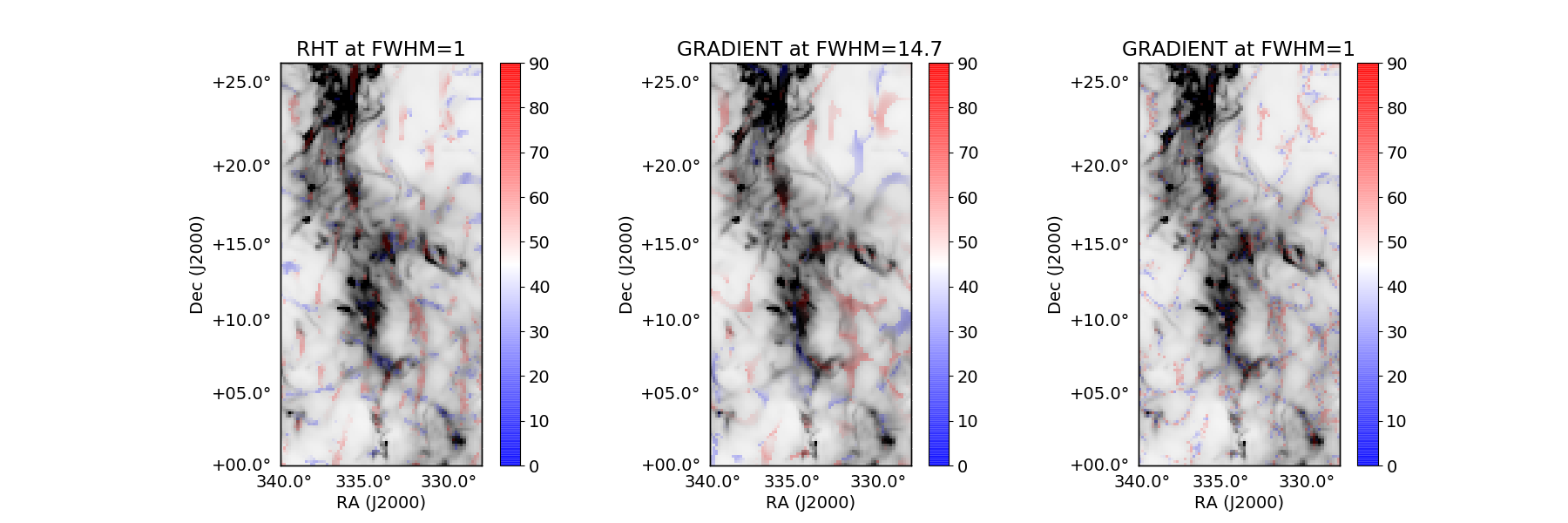} \\
    \includegraphics[width=\hsize]{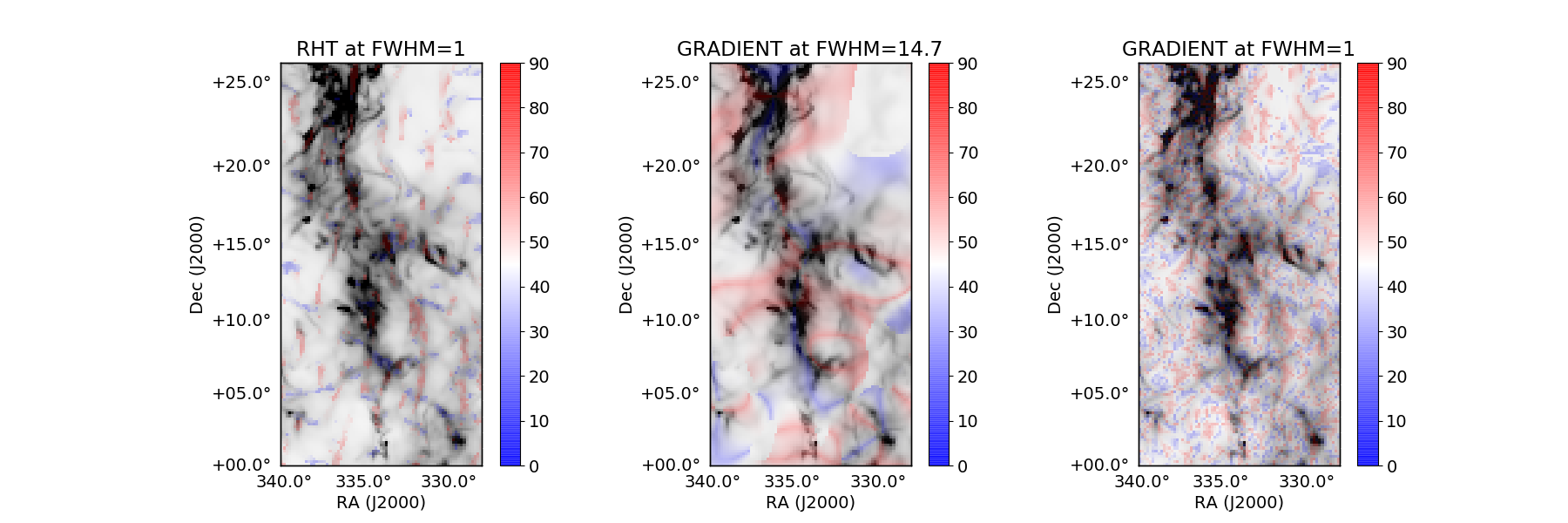} \\
  \end{center}
  \caption{Relative orientation between the magnetic field and the
    density structures (blue-red color bar) overlaid on the intensity
    map (shades of gray) kept at
    fixed resolution (FWHM=1). The relative orientation goes 
    from fully parallel (blue, 0$^{\circ}$) to fully perpendicular
    (red, 90$^{\circ}$). In the top row, the
    analysis is performed on the same pixel subsample selected by RHT
    with the significance cut only, applying the RHT method (left), 
    the gradient method smoothed to match the RHT resolution (middle)
    and the gradient method with FWHM=1 (right). In the bottom row,
    each method is performed on its own native selection.}
    \label{fig:rel_orientation_maps_reg1} 
\end{figure*}

\begin{figure*}
  \begin{center}
    \includegraphics[width=\hsize]{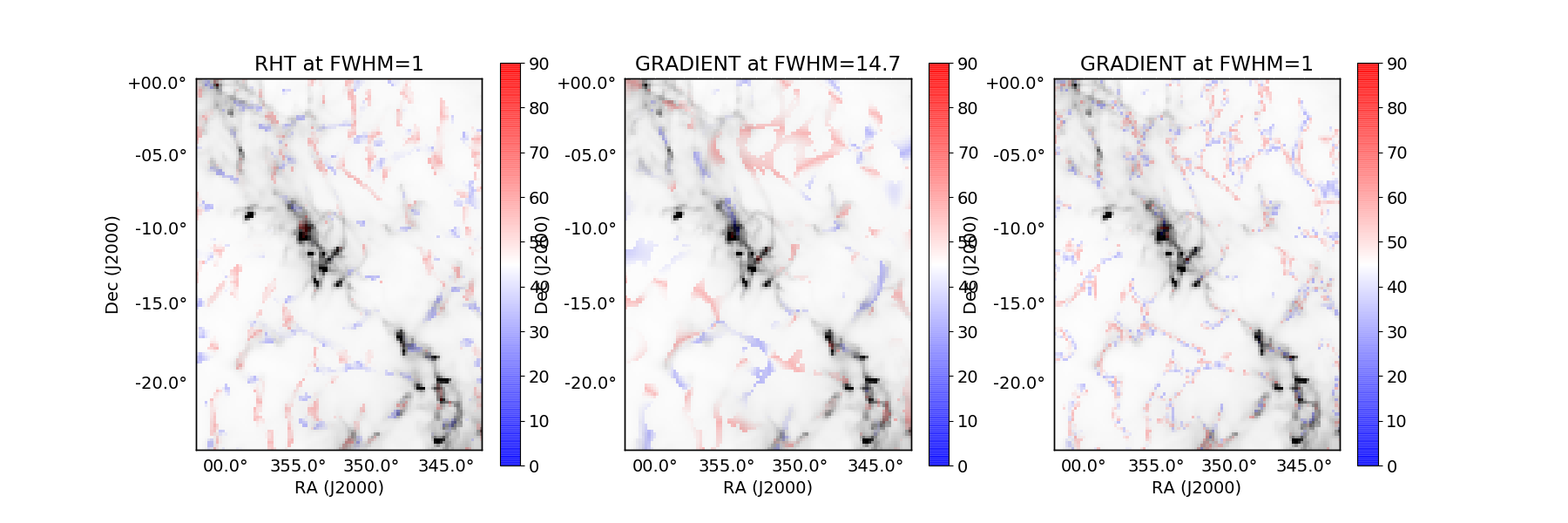} \\
    \includegraphics[width=\hsize]{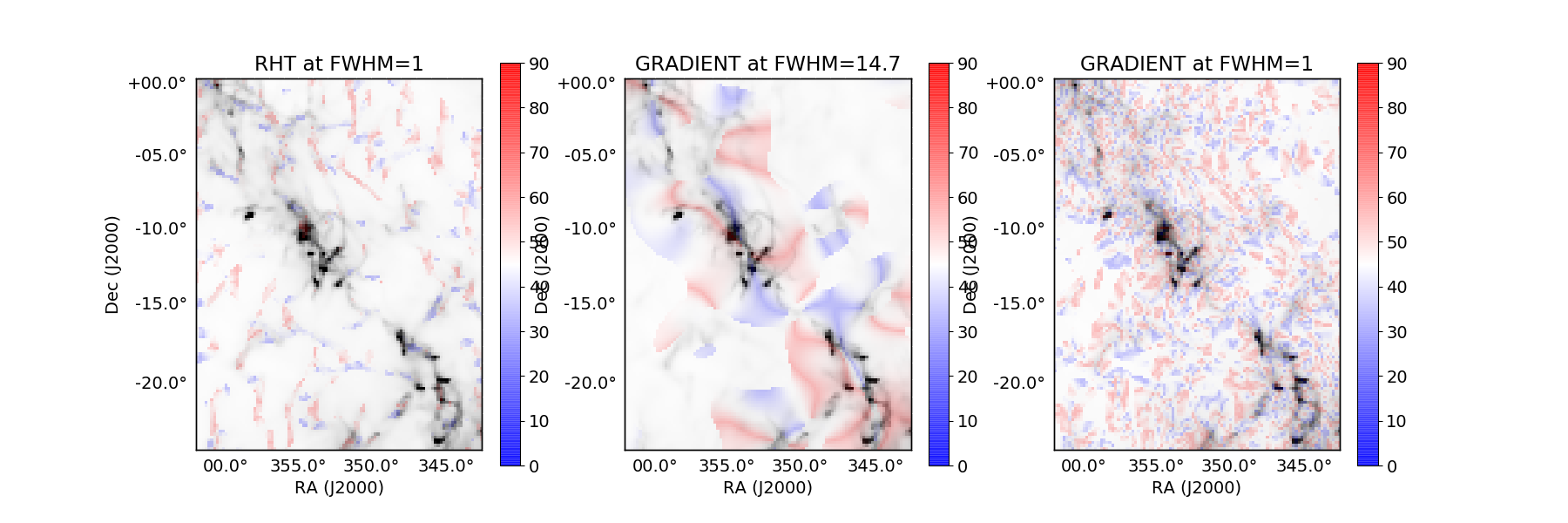} \\
  \end{center}
  \caption{Same as Fig.~\ref{fig:rel_orientation_maps_reg1} but for Region 2.}
    \label{fig:rel_orientation_maps_reg2} 
\end{figure*}

Figures \ref{fig:rel_orientation_maps_reg1} \& \ref{fig:rel_orientation_maps_reg2}  show the surface brightness maps (shades of gray) for for Region 1 and Region 2 respectively, overlaid with our analyzed pixel selection color-coded according to the relative orientation between the magnetic field and the density field. The color goes from blue for fully parallel orientation (relative angle equal to zero) to red for fully perpendicular orientation (relative angle equal to 90$^\circ$). The top row shows the results when the different methods are performed to the same pixel sample, selected by RHT with the cut on the significance only. In the bottom row, each technique has been applied to its native selection. The left column refers to the RHT method with full resolution data (FWHM=1), the central column to the gradient technique applied to data smoothed to match the resolution of RHT (FWHM=14.7) and finally, the right column refers to the gradient technique with FWHM=1 data.

In both Region 1 and Region 2, the three panels in the top row appear very similar, which confirms the fact that the analysis has been indeed performed over the same pixel selection. However, some interesting differences can be noted. For both RHT at FWHM=1 and gradient at FWHM=14.7, the colored regions are pretty smooth and have similar thicknesses, indicating that the two methods have been performed at the same resolution. On the other hand, the colored regions for the gradient at FWHM=1 (right column) exhibit a finer granularity, confirming that the method has worked at a higher resolution than in the other two cases. 

In terms of the specific orientations (the color of the regions) the RHT method and the gradient method at FWHM=1 result in a higher level of mixing, with blue and red often interwoven in the same region, while for the smoothed gradient the colored patches tend to be monochromatic, indicating only one preferred orientation in a specific spatial region. The kind of recovered orientation, preferentially parallel or orthogonal, is generally consistent in the three considered cases, with some notable exceptions from the smoothed gradient method, in particular in Region 1. For instance, in the top-right corner of the map, there are two vertical features which are blue (parallel orientation) for the gradient at FWHM=14.7 and predominantly red in the other two cases. The horizontal features between Dec +10$^\circ$ and +15$^\circ$ are red (perpendicular orientation) for the smoothed gradient and mostly blue or mixed in the other cases.

When each method is applied to its own native pixel selection (bottom row of Fig.~\ref{fig:rel_orientation_maps_reg1} and Fig.~\ref{fig:rel_orientation_maps_reg2}), the results are much more diverse. The large coverage and fine granularity given by the gradient technique at FWHM=1 (right panel) show a complex pattern where preferentially parallel and preferentially perpendicular orientations are mixed together at small scales. The RHT method (left panel) shows its capability in recovering narrow elongated structures at small scales. The blue filaments develop in the horizontal direction while the red ones develop in the vertical direction. This is consistent with the fact that in Regions 1 \& 2, the large-scale magnetic field is oriented in the $x$-direction, corresponding to the horizontal direction in the $z$-axis snapshot where the regions have been identified (right panel in Fig.~\ref{fig:polvec_combine}). The gradient method applied on smoothed data appears to be sensitive at intermediate scales. The colored patches are both extended and filamentary and the parallel (blue) and perpendicular (red) orientations alternate across the regions. In Region 1, a blue vertical filament seems to trace the backbone of the main density structure, while the horizontal branch is traced by a red filament. The main density structure in Region 2 seems also to be traced by blue strips but in a more fragmented way.

% FIGURE  *************************************************************
%
% ********************************************************************

% FIGURE  *************************************************************
%
% ********************************************************************

% FIGURE  *************************************************************
%
% ********************************************************************

% FIGURE  *************************************************************
%
% ********************************************************************

% FIGURE  *************************************************************
%
% ********************************************************************

\section{Discussion and Conclusions}\label{sec:disc_concl}

In this paper we have studied dust polarization in synthetic maps of molecular clouds generated via MHD simulations. We focused on the comparison between two methods which are widely used to analyze the relative orientations between the magnetic field, traced by polarized dust emission, and the density structures, for which dust emission properly propagated through radiative transfer is used as a proxy. The techniques that we have investigated are the RHT method and the gradient method and our goal was to establish under which conditions the comparison between the two methods gives meaningful results. We considered three specific cases: the RHT method applied to data convolved over a 1-pixel Gaussian beam (FWHM = 1), the gradient method applied to data first smoothed to FWHM = 14.7 to make it work on the same scale as the RHT technique, and finally the gradient method applied to full resolution data (FWHM = 1). 

We selected two regions in our simulated maps, named Region 1 and Region 2 and performed the relative orientations analysis in the three aforementioned cases. We first considered the same pixel selection in all cases, to ensure the same bias for all methods, and then we applied each method to its own native selection, as it is done when the analysis is performed on observed data. To quantify the behavior of the relative orientation as a function of the column density $N_{\rm H}$ we used two statistics: the HRO statistics providing  the $\xi$ parameter with uncertainty given by the Gaussian limit of the Poisson law, and the PRS giving the parameter $Z_{\rm Jow}$ with its characteristic error.

Our first finding is that the sample selection in the three cases is very different, both in terms of number and location of the selected pixels. This implies that each method will trace different regions and will be sensitive to different scales. When the methods are applied to the same pixel selection, the results in terms of $\xi$ and $Z_{\rm Jow}$ are consistent between each other (especially in Region 1) but with some noticeable differences (especially in Region 2). This implies that the output in terms of relative orientations between the density structures and the magnetic field depends on the method used to determine the relative orientation (RHT or gradient), on the resolution at which the methods are applied (FWHM = 1 or FWHM = 14.7) and on the characteristics of the analyzed regions (Region 1 and Region 2 are substantially different).

When each method is applied to its own native pixel selection, the differences are apparent. In particular, in both regions, the $\xi$ and $Z_{\rm Jow}$ curves for RHT and smoothed gradient show opposite trends for increasing values of $N_{\rm H}$. In Region 1, the RHT $\xi$ and $Z_{\rm Jow}$ curves are globally flat but start to steady decrease just after $N_{\rm H}$ = 8$\times$10$^{21}$ cm$^{-2}$. For the smoothed gradient, the behavior is almost specular and the curves tend to rise in the highest column density bins, with the transition occurring at a similar value for $N_{\rm H}$. In Region 2 the behavior is the opposite: rising trend for RHT and decreasing trend for the smoothed gradient. In both cases, the highest density bin behaves opposite with respect to the global trend. When the gradient method is applied with FWHM = 1, the trend is flat in Region 1 and rising in Region 2. In all our considered cases, the values of $\xi$ and $Z_{\rm Jow}$ are predominantly negative (perpendicular orientation), with a small fraction compatible with zero (no preferred orientation). In very few cases the values of the parameters are positive (parallel orientation), the clearest case being Region 2 with the smoothed gradient (Figs.~\ref{fig:same_pix_reg2} \& \ref{fig:diff_pix_reg2}, middle row, right panel). 

The $Z_{\rm Jow}$ curves follow the shape of the $\xi$ curves, although the absolute values of the parameters for a given column density bin can be quite different in some cases. The error bars from the PRS are always smaller than for the HRO statistics. 

As already mentioned, one of the main finding from the analysis of Planck data is that the alignment of the density structures in molecular clouds tend to be parallel to the local magnetic field, or without a preferred orientation, for low column densities ($N_{\rm H} \lesssim$ 10$^{21.7}$ cm$^{-2}$) and orthogonal to the direction of the field for high column densities \citep{soler_Planck_XXXV_16}. 
In the ten Gould Belt nearby molecular clouds analyzed in this study, the values of the histogram shape parameter $\xi$ tend to decrease from positive to negative for increasing column density, indicating that the preferred relative orientation, calculated using the gradient technique, goes from parallel to perpendicular  when moving towards denser regions. The steepness of the slope changes in the different regions. This trend has been also recovered using HROs calculated from simulations \citep{chen16}. The analysis of BLAST-Pol data of the Vela C molecular cloud \citep{soler17} has led to the same conclusions, and the observed trend has been confirmed by \citet{jow18} where both the \textit{Planck} Gould Belt and BLAST-Pol Vela C data have been re-analyzed applying the PRS. 

\citet{malinen16} analyzed Herschel data of the high-latitude molecular cloud L1642, investigating the relative orientation between magnetic field and density structures using the RHT technique. They found as well that the preferred relative orientation transitions from parallel to perpendicular to the magnetic field for increasing column density, and the transitions occurs at around $N_{\rm H}$ = 1.6$\times$10$^{21}$ cm$^{-2}$, which is equivalent to the value reported in \citet{soler_Planck_XXXV_16} when the same convention for dust opacity is used. 

\citet{alina17} performed a statistical analysis of the relative orientation in the filaments hosting the Planck Galactic Cold Clumps \citep{planck_inter_XXVIII_16}, using an improved version of the RHT method called supRHT. Their analysis revealed a change in the relative orientation between the magnetic field and the filaments which depends on both the column density of the environment and the density contrast of the filaments. The combination of low column density environment and low column density contrast results in the preferential alignment of the filaments with the background magnetic field, while low column density contrast filaments embedded in high column density environments tend to be orthogonal to the magnetic field. When the clumps inside filaments are considered, the observed alignments are both parallel and perpendicular.

When we compare our findings with the results discussed above, it appears that they are consistent \textit{in some cases}. In Region 1, we need to use the RHT method while in Region 2 the gradient technique applied at the same resolution as RHT is required. When the other method is applied (or at a different resolution) the results are different. Now the question is how to interpret these different results: should they be considered unreliable? While we consider this possibility unlikely, a more definite answer requires to perform the same analysis on observed data. We suggest that our findings reveal the intrinsic complexity of the mutual relationship between density structures and magnetic fields in molecular clouds. The RHT method and the gradient technique applied to the same patch of the sky will select and analyze different pixel sub-samples on different scales, even when they are supposed to work at the same resolution, showing different, and sometimes opposite behaviors. Locations with parallel orientation sit side-by-side with orthogonally-oriented regions (see Figs.~\ref{fig:rel_orientation_maps_reg1} \& \ref{fig:rel_orientation_maps_reg2}) indicating a complex dynamics of the density flows.

Our conclusion is that the RHT and the gradient technique applied with different resolution are complementary methods to investigate the relative orientation between density structures and the plane-of-the-sky component of the magnetic field in molecular clouds. When used together, they provide a much more complete information than each one alone, which can then be used to build a more realistic picture of the physical processes governing star-forming regions.

% % FIGURE  *************************************************************
% %
% \begin{figure*}
%   \begin{center}
%     \includegraphics[width=\hsize]{}
%   \end{center}
%   \caption{}
%     \label{fig:} 
% \end{figure*}
% % ********************************************************************

%\clearpage

\begin{acknowledgements} 
{We would like to thank D. Kodi Ramanah and G. Lavaux for providing plotting routines that were used as a basis for computing the ``drapery pattern'' in Fig.~\ref{fig:polvec_combine}. E.R.M. and M.J. wish to acknowledge the support from the Academy of Finland grant No. 285769, J.M. acknowledges the support of ERC-2015-STG No. 679852 RADFEEDBAC.}
\end{acknowledgements}

\bibliographystyle{aa}
%\bibliography{MHDPOL}

% \appendix

\end{document}